\documentclass[aps, pra,reprint,twocolumn, groupedaddress, floatfix]{revtex4-1}

\pdfoutput=1 
\usepackage{graphicx}
\usepackage{textcomp}
\usepackage{amsmath}
\usepackage{dcolumn} 
\usepackage{bm}      
\usepackage{color}

\newcommand{\bket}[1]{\left<#1\right>}

\newcommand{\ket}[1]{\left|#1\right>}
\newcommand{\bra}[1]{\left<#1\right|}

\newcommand{\RNum}[1]{\uppercase\expandafter{\romannumeral #1\relax}}

\newcommand*{\citen}[1]{%
  \begingroup
    \romannumeral-`\x 
    \setcitestyle{numbers}%
    \cite{#1}%
  \endgroup   
}

\makeatletter
 \begingroup
  \catcode`\_=\active
  \protected\gdef_{\@ifnextchar|\subtextup\sb}
 \endgroup
\def\subtextup|#1|{\sb{\textup{#1}}}
\AtBeginDocument{\catcode`\_=12 \mathcode`\_=32768}
\makeatother

\begin{document}

\title{Planar multilayer circuit quantum electrodynamics} 
\author{Z.K. Minev}
\author{K. Serniak} 
\author{I.M. Pop}
\author{Z. Leghtas}
\author{K. Sliwa}
\author{M. Hatridge}
\author{L. Frunzio}
\author{R.J. Schoelkopf}
\author{M.H. Devoret}
\affiliation{Department of Applied Physics, Yale University, New Haven, Connecticut 06511, USA}

\date{\today}

\begin{abstract} 
Experimental quantum information processing with superconducting circuits is rapidly advancing, driven by innovation in two classes of devices, one involving planar micro-fabricated (2D) resonators, and the other involving machined three-dimensional (3D) cavities.
We demonstrate that circuit quantum electrodynamics can be implemented in a multilayer superconducting structure that combines 2D and 3D advantages. 
We employ standard micro-fabrication techniques to pattern each layer, and rely on a vacuum gap between the layers to store the electromagnetic energy. 
Planar qubits are lithographically defined as an aperture in a conducting boundary of the resonators. 
We demonstrate the aperture concept by implementing an integrated, two cavity-modes, one transmon-qubit system. 
\end{abstract}
\maketitle

\section{Introduction}
Circuit quantum electrodynamics (cQED)~\cite{Blais2004, Wallraff2004}, based on the interactions of superconducting qubits with microwave light, is currently emerging as one of the most promising  experimental platforms for quantum information processing~\cite{Devoret2013, Barends2014} and quantum optics experiments~\cite{Hofheinz2009, Kirchmair2013, Bretheau2015, Leghtas2015}.
In these superconducting circuits, Josephson junctions provide the non-linearity for qubits, while low-loss microwave resonators provide linear processing functions for quantum memories~\cite{Mariantoni2011, Mirrahimi2014,Reagor2013,Vlastakis2015}, readout or entanglement buses~\cite{Majer2007, Sillanpaa2007}, and filtering~\cite{Houck2008,Reed2010}. 

Circuit QED has been developed in two platforms: fully planar  (2D) circuits, which benefit from the geometric precision and parallel production of established micro-fabrication technologies, and  3D circuits involving conventionally machined cavities, but with superior coherence times.
Through improved design and material optimization, 2D qubits and resonators have significantly progressed, with internal quality factors ($Q_|i|$) exceeding $10^6$ (Refs.~\citen{Barends2010, Megrant2012, Barends2013,Pop2014, Bruno2015}).
On the other hand, 3D resonators store a larger fraction of their electromagnetic energy in vacuum, making them less susceptible to material imperfections, and can reach $Q_|i|~>~10^8$ (Refs.~\citen{Paik2011, Reagor2013}).
Can a new cQED design take advantage of the benefits of both 2D and 3D platforms?

We propose to lithographically pattern qubits and resonators in multiple planes separated by vacuum gaps used to store the electromagnetic energy.
Thin-film aluminum resonators built in this multilayer planar way recently demonstrated low-losses ($Q_|i|~>~3\times10^6$) at the single photon level~\cite{Minev2013}. 
One of the main challenges in the implementation of a multilayer approach to cQED is the design of qubit-resonator coupling between different layers of a structure.

In the 2D and 3D platforms, coupling is achieved by inserting the qubit metallic structure onto the insulating region of the resonator. 
In a perturbative description of the coupling, the electric field of the resonator mode is aligned with the electric dipole of the qubit mode (Fig.~1a). 
However, in a multilayer architecture, this method would require fabricating qubits perpendicular to the lithographic planes.
We propose a different design strategy in which the qubit design layer coincides with one of the lithographic planes.
It uses out-of-plane fields to couple the qubit\textemdash which we nickname aperture transmon\textemdash to the resonator mode (Fig.~1b).

To demonstrate the feasibility and advantages of our multilayer planar platform for cQED, we present in this letter the implementation and coherence properties of an integrated system composed of two standing  modes coupled to a qubit, a now standard configuration for many basic cQED experiments~
\cite{Johnson2010, Kirchmair2013, Vlastakis2013, Leghtas2015, Flurin2015}. 
Our implementation is based on the two TEM modes of a superconducting whispering gallery (WG) resonator introduced in Ref.~\citen{Minev2013}.
One of the modes is over-coupled ($Q=10^4$) to a readout amplification chain, while the other is maintained as high-Q as possible ($Q=2\times 10^6$).
Both modes couple to a transmon qubit~\cite{Koch2007} with a $T_1 = 70\ \mu$s lifetime. 
The Hamiltonian of this device (see Eq.~\ref{eqSysHamiltonian}) is similar to that of the 3D device in Ref.~\citen{Kirchmair2013}, and can be used for the implementation of cavity-based error correction protocols~\citen{Mirrahimi2014}.


\section{Device and Methods}

Figure~\ref{fig:Coupling} shows the multilayer chip-stack elements of the measured device.
Two two sapphire chips served as substrates for each of the two Al patterned rings.
We positioned the chips with the rings aligned inside an Al sample holder to establish the boundary conditions of the TEM modes of the WG resonator.
Machined ledges in the sample holder maintained a 100~$\mu$m vacuum gap between the chips (details of the assembly can be found in the Appendix).
The two orders of magnitude in aspect ratio between the mode wavelength and the stack gap-spacing ensured tight confinement of the fields of the modes.
In particular, the inductive participation ratio of the sample holder was found to be $10^{-8}$ or smaller for each of the modes, as computed with an HFSS~\footnote{High frequency structural simulator (HFSS) from ANSYS, Inc.} finite element model, assuming a London penetration depth of $50$~nm.

\begin{figure}[t]
\begin{centering}
\includegraphics[width=3.375in]{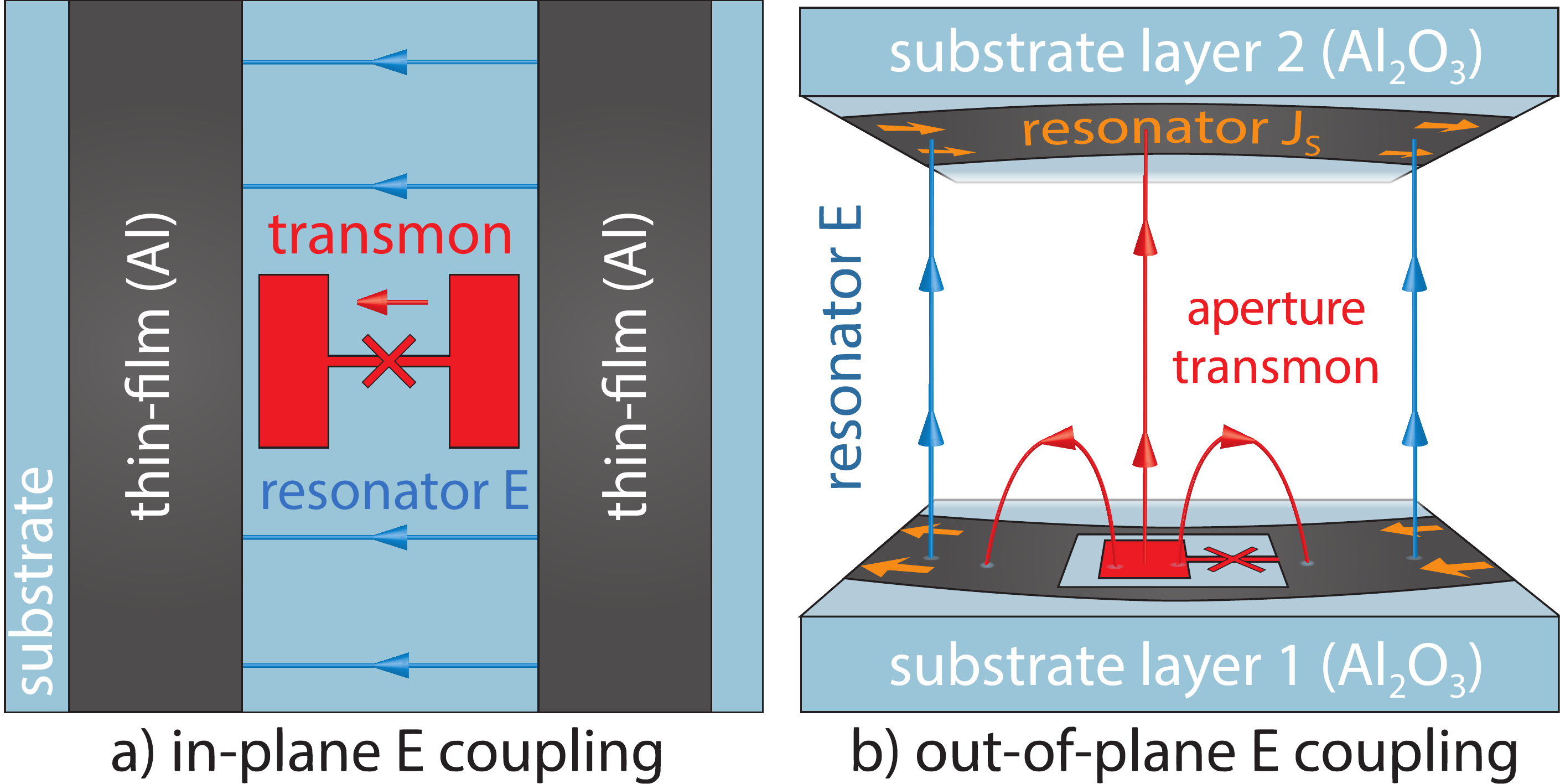}
\caption{
Qubit-resonator coupling in different cQED approaches. 
a) 
In-plane coupling in 2D. 
The electric field lines of the resonator (blue) are aligned with the dipole moment of the qubit (red), both of which are in the plane of qubit fabrication.
b)
Out-of-plane coupling in a multilayer planar device. 
The resonator is now represented as a section of a multilayer whispering gallery mode resonator~\cite{Minev2013}, consisting of two superconducting thin-film rings deposited on different sapphire 	substrates that are separated by an electrically-thin vacuum gap. 
The qubit is defined by an aperture carved directly from the conducting boundary of the resonator.
The orange and blue arrows represent the resonator surface current density and electric field lines, respectively.
}
\label{fig:PlnarNonPlanarCouple}
\end{centering}
\end{figure}

The transmon qubit was directly patterned in the thin-film of the ring in layer 1, as shown in the inset of Fig.~2. 
The qubit consists of a $0.05 \times 0.5$~mm island inside a $0.23 \times 1$~mm aperture in the ring, connected through a Josephson junction with $E_|J| /h = 12$~GHz. 
The qubit structure perturbs the resonator mode frequencies only at the percent level.
The island nominally shares 60~fF of capacitance with its own ring and 5~fF with the opposite ring. 
The junction capacitance and these geometric capacitances define the nominal qubit charging energy $E_|C|/h = 275$~MHz, frequency $\omega_|q|/2\pi = 4.85$~GHz, anharmonicity $\alpha = 320$~MHz~\cite{Koch2007}.
The rings and the qubit are fabricated simultaneously in a single electron-beam lithography step using a double-angle, bridge-free technique~\cite{Lecocq2011, Rigetti2009}.

The spatial mode orthogonality of the two WG modes allowed us to implement the long lived storage ($D^\perp$) and over-coupled readout ($D^\parallel$) modes of a quantum register within the same physical structure.   
In the following, we refer to these two modes simply as ``storage"  and ``readout," with nominal coupling $Q_|c|^\textrm{S} > 10^8$ and $Q_|c|^\textrm{R} = 1.8\times 10^4$ (see Appendix~\ref{sec:IOcoupling}), respectively. 


\begin{figure}[!t]
\begin{centering}
\includegraphics[width=3.375in]{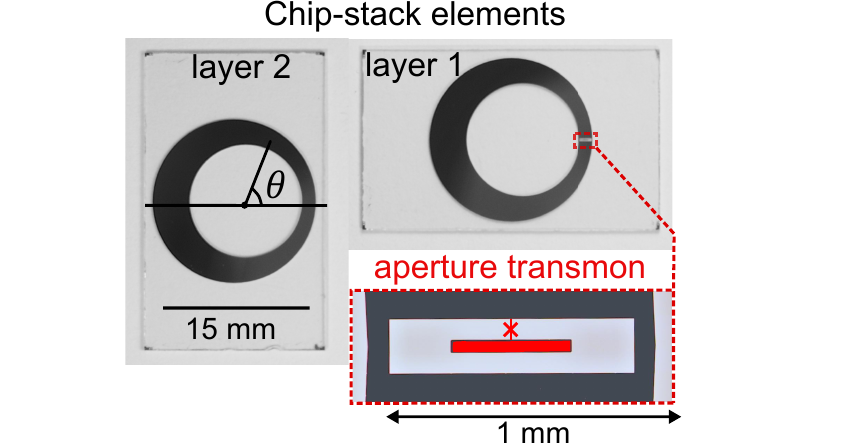}
\caption{
Photograph of chip-stack elements.  
Thin-film Al rings are patterned in a single e-beam lithography step along with the Josephson junction on a sapphire substrate. 
The boxed region shows a magnified optical micrograph of the embedded aperture qubit (false-colored in red) and the location of the Josephson junction (red cross).
The axis of symmetry, represented over  substrate layer 2, defines the angular position $\theta$ around the ring.  
}
\label{fig:Coupling}
\end{centering}
\end{figure}

\begin{figure*}[t]
\begin{centering}
\includegraphics[width=6.9in]{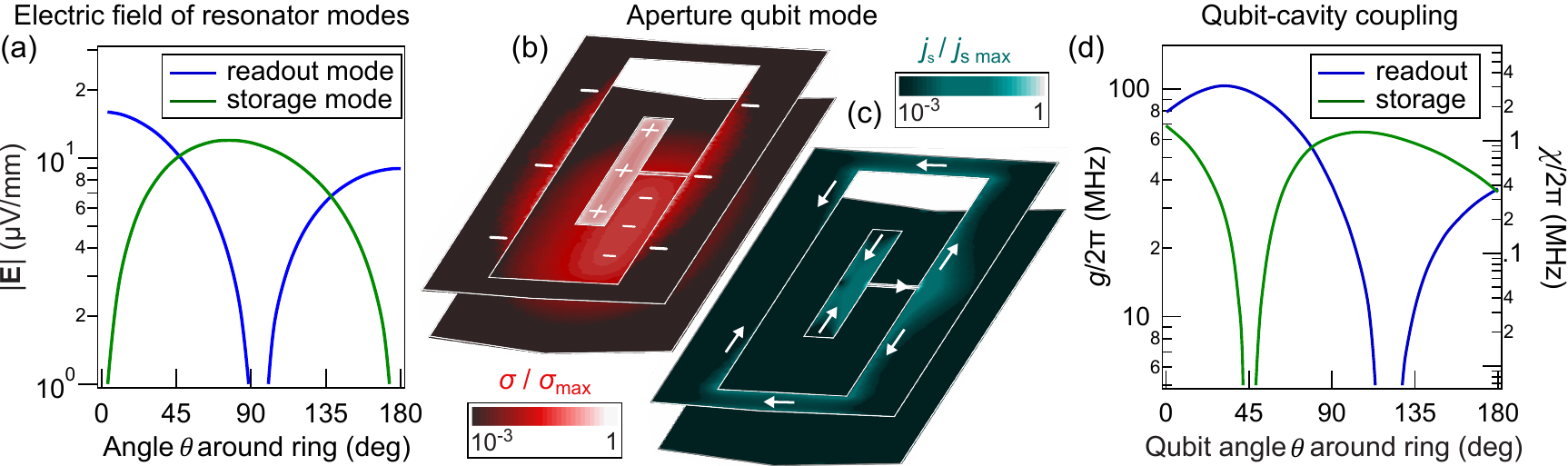}
\caption{
Electric and magnetic contributions to qubit-resonator coupling, as calculated using HFSS.
a)
Electric field amplitude $|\mathbf{E}|$ in the resonator WG modes as a function of $\theta$, for one photon of energy.
The angular dependence of the surface-current-density  $j_|s|$  follows that of $E$, but is shifted by $90^{\circ}$.
b)
Surface-charge-density amplitude $\sigma$ shown in color scale with overlayed white signs to indicate the relative charge polarity.
The charges in the island and corresponding image charges in the opposite layer below determine the electric contribution to the qubit-resonator coupling.
c) 
Qubit mode surface-current amplitude  $j_|s|$ shown in color scale with overlayed white arrows to represent the direction of the flow.
The narrow rails on each side of the aperture are equivalent to a shared inductance between the qubit and resonator and determine the magnetic contribution to the qubit-resonator coupling.
d) 
Qubit-resonator coupling rate $g$ of the aperture qubit to the WG resonator modes as a function of the qubit position around the ring $\theta$. 
The right vertical axis also shows the equivalent cross-Kerr~$\chi$ for a $\Delta=1$~GHz detuning.
The coupling $g$ is the algebraic sum of the electric and magnetic contributions, which interfere constructively or destructively as a function of $\theta$.
For the readout (storage), maximal constructive (destructive) interference occurs at about $45^{\circ}$, while near $\theta=0, 180^{\circ}$ the coupling is capacitive (inductive).  }
\label{fig:figEandChi}
\end{centering}
\end{figure*}

The sample holder was thermally anchored to the 15~mK stage of a dilution unit. 
We used the standard cQED measurement setup (See Appendix~\ref{sec:Experimental}) with the addition of a phase preserving, quantum limited, Josephson parametric amplifier~\cite{Bergeal2010}.

\section{Planar to non-planar coupling}
For dispersive coupling between a transmon qubit (q) and a resonator mode (r), the strength of the cross-Kerr  $\chi_{qr}$ depends primarily on the detuning $\Delta=\omega_|r|-\omega_|q|$, aperture geometry, and the resonator fields at its position. 
In order to quantify the aperture coupling independently of the potentially tunable $\Delta$, we define the effective coupling rate $g_|qr| = \Delta\sqrt{\chi_|qr|(\Delta)/E_|C|}$~\citen{Koch2007}, which is approximately independent of $\Delta$ and $E_|J|$ in the transmon limit $E_|J| /E_|C| \gg 1 $ and for weak interaction $g\ll \Delta$.

In 2D and 3D, the coupling strength $g$ can be understood as arising from an  interaction between the bare electric field of the resonator and the electric-dipole-like charge distribution of the qubit.
Here, in our multilayer structure where the qubit is patterned in an aperture in one of the layers, the coupling mechanism is more involved.
Figure~\ref{fig:figEandChi}b and ~\ref{fig:figEandChi}c show the charge and current distribution of the qubit mode, respectively.
The coupling is determined by the overlap between these distributions and those of the resonator mode.

The interplay of the capacitive (charge overlap) and inductive (current overlap) coupling is shown in Fig.~\ref{fig:figEandChi}d, where we plot the dependence of $g$ on the qubit position $\theta$ (see Appendix ~\ref{sec:SimulMethodology}) for the simulation procedure based on black-box circuit quantization~\cite{Nigg2012}).
Varying the position $\theta$ of the transmon varies its coupling $g$ to the resonator, independently of its frequency and anharmonicity. 
For a given position $\theta$, $g$ can be further adjusted by changing the dimensions of the aperture.

\section{Experimental results} 

Microwave spectroscopy revealed the transmon qubit,  storage  ($D^\perp$), and readout ($D^\parallel$) modes at 4.890~GHz, 7.070~GHz, and  7.267~GHz, respectively, in~1\% agreement with the HFSS numerical simulations of the sample \footnote{This simulation included observed non-ideality in the alignment of the rings.}. 
From qubit spectroscopy, we observed a 310~MHz  transmon anharmonicity, in~5\%  agreement with the nominal qubit charging energy~\cite{Koch2007}.

\begin{figure}[t]
\includegraphics[width=3.375in]{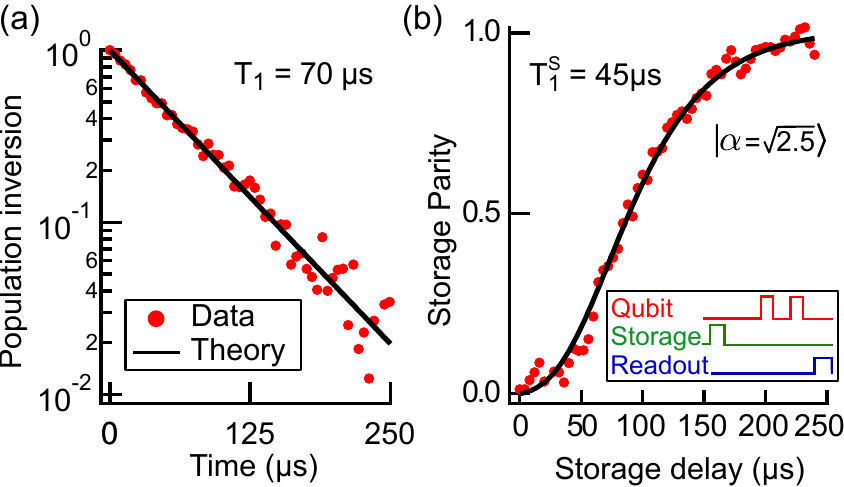}
\caption{
a) 
Aperture transmon energy relaxation. A single exponential fits the data with a $T_1 = 70~\mu$s.
The population inversion is defined as $(P_|e|-P_|e|^{\mathrm{th}} )/ (P_|g|^{\mathrm{th}} - P_\mathrm{e}^{\mathrm{th}})$, where $P_|e|^{\mathrm{th}}$ and $P_|g|^{\mathrm{th}} $ are the thermal populations. 
b) 
Storage parity relaxation from which we infer an energy relaxation lifetime of $T_1^\textrm{S}=45 \mu$s.
Inset: pulse sequence for the measurement, see Appendix~\ref{sec:calib}.}
\label{fig:StrgLifetime}
\end{figure}

Figure~\ref{fig:StrgLifetime}a shows the qubit free decay with an exponential time constant $T_1 = 70\ \mu$s. 
We measured $T_2^\mathrm{R} = 8~\mu$s and $T_2^\mathrm{E} = 20~\mu$s.
The dephasing noise was measured by a CPMG technique (see Fig.~\ref{fig:figLifetimes}).
The contributions to the dephasing noise are not currently understood, they could be the result of photon shot noise~\cite{Sears2012},  mechanical vibrations and/or offset charge drifts~\cite{Koch2007}.
The readout  linewidth, $\kappa_\mathrm{r}/2 \pi  = 0.35$~MHz, and qubit dispersive shift, $\chi_|qr|/2 \pi = 0.30$~MHz,  agreed to~10\% with simulations.

From spectroscopy of the storage at photon number $\bar{n} \approx 10^2$, we measure a linewidth  $\Delta\omega_|S| / 2\pi= 4~\text{kHz}$ which includes self-Kerr broadening in addition to dephasing.
From this linewidth, we infer an approximate lower bound on the storage coherence time: $T_2^\mathrm{S} \gtrsim  2/\Delta\omega_|S|  = 80~\mu$s.

To measure the storage lifetime $T_1^\mathrm{S}$ in the single-photon regime (see Fig.~\ref{fig:StrgLifetime}b), we used the photon-number parity protocol introduced in Ref.~\cite{Kirchmair2013, Vlastakis2013, Sun2014}, which, in our case, is more sensitive than a direct amplitude decay measurement.
A 1~$\mu$s Gaussian pulse first displaced the storage cavity to a coherent state with $\bar{n} = 2.5$~photons, a state with essentially zero parity $P_|s| = \left\langle \exp(-i \pi a^\dagger a)\right\rangle = 5 \times 10^{-3}$, where $a$ is the storage lowering operator. 
After a variable delay, a Ramsey-like sequence with a fixed time delay of $\pi/\chi_|qs|$ mapped the parity of the storage photon number to the qubit state.
From the parity measurement, we extracted a low-photon number  $T_1^\mathrm{S} = 45~\mu$s.
Measurements at higher photon numbers (up to $\bar{n} = 200$) showed no power dependence of $T_1^\mathrm{S}$.
The calibrations needed for this protocol are described in Appendix~\ref{sec:calib}.

\section{Discussion} 
We summarize the two cavity modes, one-qubit device interaction strengths and lifetimes in  Table~1.
The measured frequencies and coupling energies of the  multilayer device agree at the percent and ten-percent level, respectively, with design values from numerical simulations.
The discrepancy can be explained by machining tolerances ($25~\mu$m) of the gap spacing and chip alignment in the sample holder, and could be improved by using micro-machined separators to support the structure~\cite{Takahashi2001, Harle2003, Brecht2015}.

The measured coherence times are on par with those of two-cavity, one-qubit devices using 3D rectangular cavities~\cite{Kirchmair2013, Vlastakis2013, Sun2014, Leghtas2015, Flurin2015, Heeres2015}.
The qubit and storage lifetimes were not limited by their input-output (I/O) coupling~\ref{fig:figIOcoupling}.
These lifetimes could be extended by design optimization, as well as material advances demonstrated in the 2D and 3D architectures~\cite{Barends2010, Chang2013, Megrant2012, Reagor2013, Quintana2014, Bruno2015}.
Spurious fringing fields in the substrates and environment would be reduced by decreasing the gap spacing and improving the chip stack alignment. 
In the present device, the 100~$\mu$m gap captures~$\sim$$90\%$ of the cavity energy and~15\% of the qubit energy in the vacuum.
A decrease in the gap by a factor of~10 would, for both modes, decrease the bulk dielectric participation down to the percent level, a gain of more than one order of magnitude over planar and 3D qubits, provided that we would not be limited by the surface quality of the superconducting film.

We demonstrated at least three orders of magnitude separation in I/O coupling Q between the storage and readout, which not only share the same physical footprint, but differ by only 200~MHz in frequency.
This type of spatial mode control is advantageous when dealing with cross-talk and frequency crowding in devices with increased complexity.

The measured  device is a suitable candidate for a quantum register~\cite{ Leghtas2013, Mirrahimi2014}, with storage coherence time $T_2^{\mathrm{S}} = 80\ \mu$s  exceeding that of the qubit by an order of magnitude. 
The storage could provide a large Hilbert space to encode quantum information, while the long-lived qubit serves as a conditional, non-linear control over the cavity space with a low bit-flip error rate. 

\begin{table}[t]
 \begin{center}
\setlength\extrarowheight{3pt}
   \begin{tabular}{l | ccc} 

    Mode & Qubit & Storage & Readout\\
   \hline\hline
   Frequency (GHz) & 4.890 & 7.070 & 7.267\\
   \hline
   $T_1$ ($\mu$s) & 70 & 45 & 0.42\\
  \hline
   $T_2 \mid T_2^{\mathrm{E}}$ ($\mu$s) & $8^\star \mid 20$ &  $\gtrsim80$ & - \\
  \hline
   $\alpha/ 2 \pi$ (MHz) & 310  &$1 \times 10^{-4}$&$1.5\times 10^{-4}$\\
    \hline		
    $\chi_|q|/ 2 \pi$ (MHz)&- & 0.25 & 0.30\\
   \end{tabular}
 \caption{
Main parameters of sample.
The cross-Kerr interaction with the qubit mode  is denoted $\chi_|q|$, while $\alpha$ denotes anharmonicity.
All parameters are measured except the storage/readout anharmonicity, which are calculated from $\alpha_|s,r| = \chi_|q\ s,r|^2/4\alpha_|q|$~\cite{Nigg2012}.
The symbol $^\star$ indicates a Gaussian decay. } 
  \end{center}
\end{table}

The qubit-resonator coupling geometry presented in Fig.~1b relies on the use of an aperture in one of the metal layers.   
Radiation fields from an aperture usually constitute spurious loss and cross-talk mechanisms, but in our case, this effect is mitigated by the proximity of the opposite superconducting layer.
In fact, our work demonstrates that we can put these fields to a good use: mediating the coupling between the  planar qubit and multilayer resonator.
This approach can be extended to provide low cross-talk inter-layer connections for devices with more than two layers, such as the architecture proposed in Ref.~\cite{Brecht2015}.

\section{Conclusions and perspectives} 
We have implemented a multilayer superconducting device for quantum information processing that combines the benefits inherent to the precise geometry control of 2D micro-fabrication with those of the coherence in 3D qubits and resonators.
In particular, the qubit-resonator mode couplings can be precisely adjusted.
We believe that the quality of the measured coherence in the present work resulted from the confinement of electric fields within the vacuum gap separating lithographically defined layers. 
The  design principles  illustrated by our work can be extended to devices with more than two layers,  each layer corresponding to a specific function: qubits, control lines, resonators, amplifiers, etc. 
In particular, the aperture based coupling method introduced here can be generalized to inter-layer coupling in such multilayer devices.
Furthermore, the TEM mode structure and the separation of layers provides a favorable geometry for hybrid systems, such as spin-ensembles with cavities~\cite{Julsgaard2013, Putz2014}, spin qubits with magnetic contacts~\cite{Cottet2010}, or nano-wire qubits~\cite{Larsen2015, DeLange2015}. 
\section{Acknowledgements}
We thank T. Brecht,  M. Reagor, C. Wang, S. Shankar, M. Rooks, for valuable discussions. 
Facilities use was supported by YINQE and NSF MRSEC DMR 1119826. 
This research was supported by ARO under Grants No. W911NF-14-1-0011 and N0014-14-1-0338, and ONR under Grant No. N0014-14-1-0338.

\appendix
\renewcommand{\thefigure}{A\arabic{figure}}
\setcounter{figure}{0}

\section{Design details} 

\textbf{System Hamiltonian.}
If we expand the Josephson junction cosine potential to fourth order, apply the rotating wave approximation, and limit the Hilbert space of the transmon mode to the first two levels~		\cite{Nigg2012}, then the effective device Hamiltonian is:
 \begin{equation} \label{eqSysHamiltonian}
 \begin{split}
 \mathcal{H}/\hbar =&  \frac{\omega_|q|}{2}(1+\sigma_|z|) + \omega_|s| a^\dagger a + \omega_r b^\dagger b \\
 &- \frac{1}{2}  (1+\sigma_|z|) (\chi_|qs|a^{\dagger} a  + \chi_|qr| b^{\dagger} b)-  \chi_|sr| a^{\dagger} a b^{\dagger} b\, ,
 \end{split}
 \end{equation}
where  $a$ and $b$ are storage and readout bosonic operators, respectively, and  $\sigma_|z|$ is a qubit Pauli operator.
The storage-readout cross-Kerr  is $\chi_|sr| \approx \chi_|qs| \chi_|qr| / \alpha_|q| $, where $\alpha_|q|$ is the transmon anharmonicity.

\begin{figure}[h!]
\includegraphics[width=3.375in]{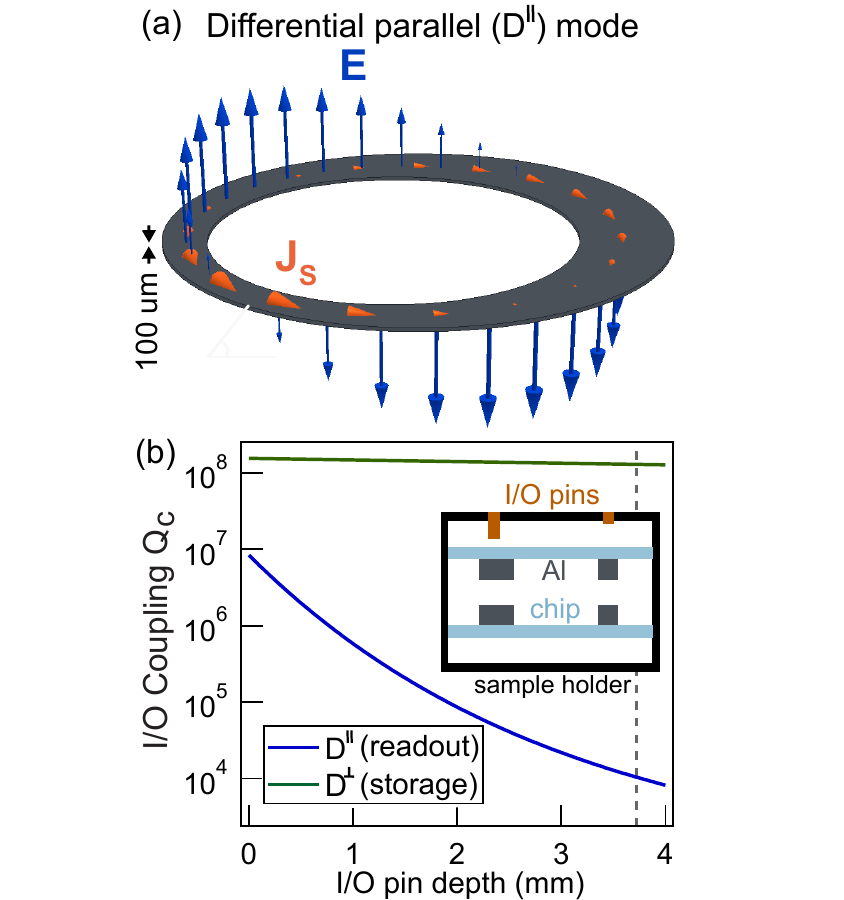}
\caption{
a)
HFSS calculation of the surface currents and electric fields, which are contained within the $100~\mu$m vacuum gap separating the two layers, for the TEM $D^\parallel$ WG mode. 
The orthogonal $D^\perp$ mode closely resembles the $D^\parallel$, up to a $90^{\circ}$ rotation.
Depending on the configuration of the input/output (I/O) coupling pins (see c) these modes can be used either as a qubit readout mode (low coupling-quality-factor $Q_|c|$) or storage for quantum states (high $Q_|c|$).
b)
Simulated $Q_|c|$ as a function of the coupling coax pin depth inside the sample holder at $\theta = 0^{\circ}$. 
The inset shows a not-to-scale cross-sectional representation of the chip stack in a sample holder (black).
We can selectively couple to $D^\parallel$ ($Q_|c|^\mathrm{R} = 10^4$) while remaining uncoupled from $D^\perp$ ($Q_|c|^\mathrm{S}  > 10^8$), as indicated by the vertical gray line, which corresponds to the nominal parameters of the measured device.
}
\label{fig:figIOcoupling}
\end{figure}

\textbf{I/O coupling.} \label{sec:IOcoupling}  
Figure~\ref{fig:figIOcoupling}a shows a simulated field profile for the $D^\parallel$ mode, where the maximum currents flow parallel to the axis of symmetry of the rings~\cite{Minev2013}.
The orthogonal $D^\perp$ mode corresponds to exchanging the field and current maxima and minima. 

As illustrated in the inset of Fig.~\ref{fig:figIOcoupling}b, two non-magnetic pins penetrated the sample holder lid to couple capacitively with the maximal charge densities of the readout  above the thinnest ($\theta = 0$) and thickest ($\theta = 180^\circ$) parts of the rings.
Owing to the selective coupling due to the spatial mode orthogonality, the nominal readout coupling $Q_|c|^\mathrm{R}$ was  $1.8\times 10^4$, while the nominal storage mode coupling $Q_|c|^\mathrm{S}$ exceeded $10^8$.

\section{Experimental details}  \label{sec:Experimental}
\textbf{Fabrication.}
We micro-fabricated both layers on the same 430~$\mu$m thick, double-side-polished, c-plane sapphire wafer.
Using a 100~kV Vistec electron beam pattern generator, we defined the WG resonator and qubit in a single lithography step on a PMMA/MAA resist bilayer.
We then performed a double angle Al evaporation, 20 and 30~nm thick, in a Plassys UMS300 at a pressure of $5\times 10^{-8}$~Torr.
Between these two depositions, an AlOx barrier was formed by thermal oxidation for 6 minutes in a static 100~Torr environment of~85\% argon and~15\% oxygen. Chips were diced to $15.5 \times 25.4$~mm.

\textbf{Qubit design details.}  
The qubit island inside is connected to the ring by a $1~\mu$m thin wire and a $130 \times 700$~nm Josephson junction with $E_J /h = 12$~GHz.
The value $E_|J|/E_|C| = 44$ yields a maximum offset-charge dispersion of $30$~kHz. 

\textbf{Sample holder.}
The chips are placed inside the bottom piece on ledges that are machined in the Al walls. 
The sample holder top piece has four legs which use indium to secure the chips against the sample holder bottom. 


\textbf{Measurement setup.}
An aluminium and permalloy shield protected the sample from stray magnetic fields. 
The SMA input lines had thermalized cryogenic attenuators (20,10,30~dB) at the 4~K, 0.1~K and 15~mK stages of a Cryoconcept DR-JT-S-200-10 dilution refrigerator, respectively. 
The sample holder output connected to a Josephson parametric converter (JPC) amplifier~\cite{Bergeal2010} through two Pamtech 4-8 GHz cryogenic circulators and superconducting NbTi-NbTi coax cables.
The JPC served as a phase preserving amplifier which operated near the quantum limit with a gain of 21~dB over a bandwidth of 5.6~MHz.
Two 4-8~GHz circulators together with two low-pass filters---a 12~GHz K\&L multi-section lowpass and a box-type Eccosorb CR-110 filter---serve to isolate the JPC from the following Low Noise Factory HEMT with 40~dB of gain. 
We found an 8 dB noise rise figure for the amplification chain, indicating that the observed noise at room temperature is~$\sim$$90\%$ percent amplified quantum fluctuations, though the total quantum efficiency of the measurements is lower due to losses. 
At room temperature, a 30~dB Miteq amplifier further amplifies the signal and feeds it into a standard heterodyne microwave interferometer operating at an intermediate frequency of 50~MHz. 
An analog to digital converter records the mixed-down output signal together with a mixed-down reference of the input signal. 
Combining the output and reference signals accounts for any phase drift in the readout control generator. 

At room temperature, we used a Tektronix~5014C arbitrary waveform generator, an Agilent~E8257D vector generator, and several Vaunix Lab Brick generators to generate the qubit and cavity tones.

\section{Details of qubit and storage mode coherence measurements.}

The qubit parameters were in the convenient regime for continuous state monitoring, where the dispersive shift and output coupling rate are nearly equal.
From quantum jump measurements, not presented here, we inferred a qubit excited state population below~4\%.

\begin{figure}[th!]
\includegraphics[width=3.375in]{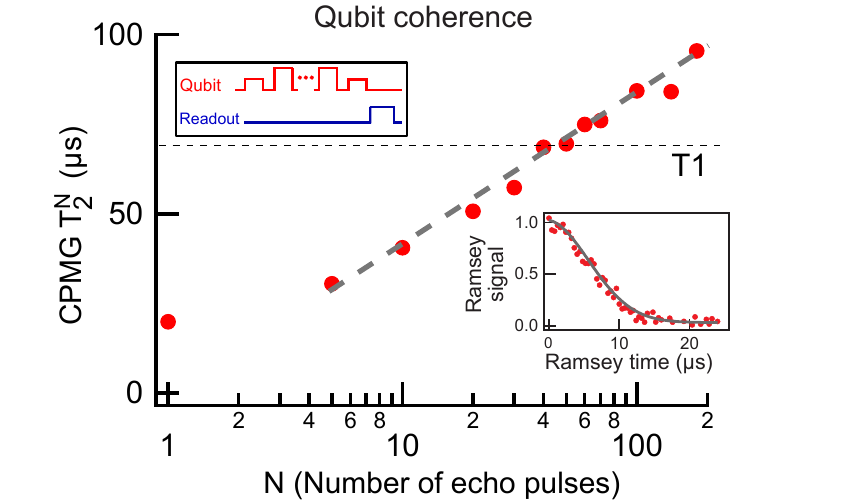}
\caption{
We measured $T_2^\mathrm{R} = 8~\mu$s (inset) and $T_2^\mathrm{E} = 20~\mu$s. 
By dynamically decoupling the qubit from low-frequency noise we observed an order of magnitude improvement in the coherence time of the aperture transmon exceeding $T_1$.  
The dashed line is a guide for the eye.   }
\label{fig:figLifetimes}
\end{figure}

\textbf{Qubit coherence.}
The inset of Fig.~\ref{fig:figLifetimes} shows the Ramsey coherence signal of the qubit which decays with a Gaussian envelope and a time constant $T_2^\mathrm{R} = 8\ \mu$s.
Since this Gaussian envelope is indicative of low-frequency noise, we used dynamical decoupling techniques to access the intrinsic qubit coherence.
A  Carr-Purcell-Meiboom-Gill (CPMG) protocol, following the approach and pulse-train calibrations of Ref.~\cite{Bylander2011}, shifted the maximum of the longitudinal noise-susceptibility of the qubit to higher frequencies. 
Figure~\ref{fig:figLifetimes} shows  the increase of the dynamically decoupled coherence time $T_2^\mathrm{N}$ as a function of the number of CPMG pulses.
The improvement of  $T_2^\mathrm{N}$ beyond $T_1$ confirms the dominance of low-frequency noise. 
The Gaussian envelope, indicative of low-frequency noise, could be the result of  mechanical vibrations and/or offset charge drifts.

\begin{figure}[th!]
\includegraphics[width=3.375in]{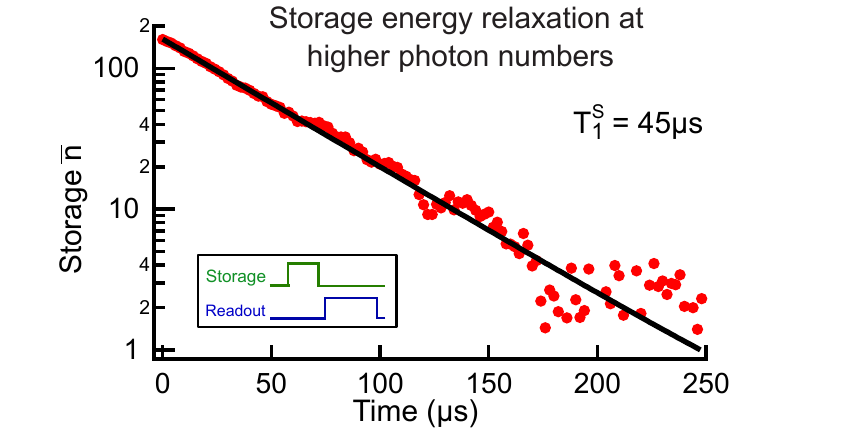}
\caption{
Using the dispersive cross-Kerr interaction with the readout mode $\chi_|sr|/2\pi= 0.25$~kHz (pulse sequence in inset), we measured a storage lifetime $T_1^\mathrm{S} = 45~\mu$s for $\bar{n} > 10$ photons.  }
\label{fig:AStrgLifetime}
\end{figure}

\textbf{Storage mode lifetime.}
To measure the storage $T_1^\mathrm{S}$ at the 10-200 photon level, we used a dispersive readout of the storage ring-down (see Fig.~\ref{fig:AStrgLifetime}), which, in our case, is more sensitive than a direct amplitude decay measurement. 
The dispersive readout is based on the  cross-Kerr frequency shift of the readout mode due to the storage photon occupation.
We apply a  500~$\mu$s coherent pulse to excite the storage mode, followed by a 250~$\mu$s, low-power tone to probe the readout frequency.
The storage photon population decayed exponentially with a lifetime $T_1^\mathrm{S} = 45~\mu$s.
The resolution of this dispersive measurement is too low to access the single-photon regime, because  of the small ratio ($10^{-3}$) between the cross-Kerr and the readout linewidth.

\textbf{I/O coupling.} 
Over several cooldowns, we progressively  decreased the I/O coupling by shortening the coupling pin lengths.
However, while the over-coupled readout  lifetime increased by a factor of two to the value in Table~1, the qubit and storage mode lifetimes, as well as the frequencies and non-linear coupling strengths, did not change measurably. 
From this, we place a lower bound on the measured storage coupling $Q_|c|^\mathrm{S} > 10^8$.

\section{Photon number parity calibration with qubit-state revivals}  \label{sec:calib}

\begin{figure}[ht!]
\includegraphics[width=3.375in]{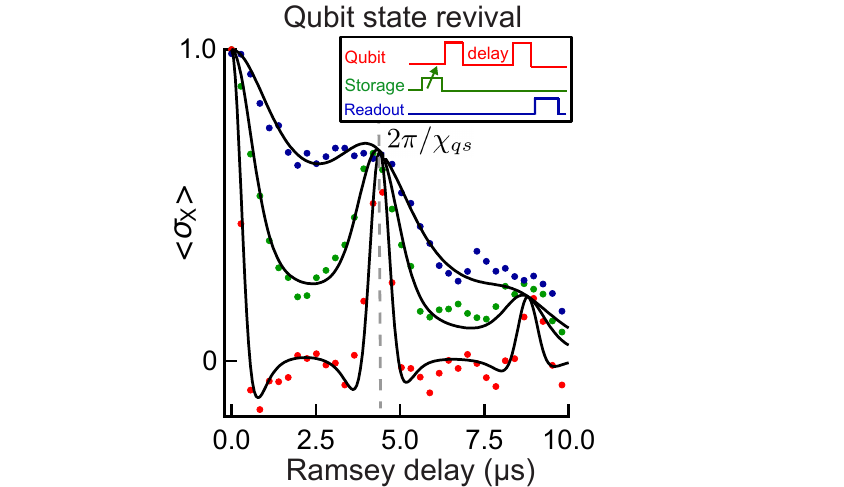}
\caption{
Calibration of the photon number parity measurement in the storage was achieved with a qubit-state revival experiment.
For small storage mode displacements  $\bar{n}_|s| \sim 0$  (blue), the decay is dominated by intrinsic qubit decoherence. 
For increasing displacements, up to $\bar{n}_|s| = 3$ (red), the apparent increase in decoherence is due to the large qubit-cavity interaction rate, and we observe qubit-state revivals at integer multiples of $2 t_|p| = 2\pi/\chi_|qs|$. }
\label{fig:figParityCalib}
\end{figure}

\textbf{Experimental method of parity measurement and calibration.}
To measure the storage photon-number parity presented in  Fig.~4b, we perform a protocol introduced in Refs.~\citen{Vlastakis2013, Sun2014}.
After displacing the storage using a coherent drive, we apply a $\pi/2$ pulse, which activates the qubit-storage cross-Kerr interaction and fully entangles the storage parity with the qubit at a time  $t_|p|$ after the qubit pulse.
A second $\pi/2$ pulse maps the parity to the expectation value of the qubit $\sigma_|z|$ operator, which is read out projectively. 
This is the measurement sequence used for the parity measurement of Fig.~4b.

This protocol requires calibration of the initial displacement photon-number $\bar{n}$ and the parity mapping delay time $t_|p| = \pi/\chi_|qs|$.
To perform the calibration, we displaced the WG storage by a short, coherent drive, and then performed a standard qubit $T_2$ Ramsey experiment, as shown in  Fig.~\ref{fig:figParityCalib}. 
Sharp coherence peaks stroboscopically reappear at integer multiples of the cross-Kerr interaction period $2 \pi / \chi_|qs|$, indicating the value of $\chi_|qs|/2\pi= 0.25$ MHz.
From a global fit to the theory (Eq.~\ref{eqQubitStateRevival}) over all displacement amplitudes, we  calibrate the corresponding  storage photon numbers $\bar{n}$.

\textbf{Calibration theory.} 
During the measurement, the readout mode remains unpopulated, and we can ignore its contribution to the system Hamiltonian from Eq.~\ref{eqSysHamiltonian}. 
In the rotating frame of the storage and qubit, the system Hamiltonian takes the form:
$$ H/\hbar =  -\chi_|qs| a^\dagger a  \ket{e}\bra{e}\, .$$
For a system starting in the ground state, the calibration Ramsey signal of Fig.~\ref{fig:figParityCalib}  obeys the following form as a function of time $t$: 
\begin{equation} \label{eqQubitStateRevival}
\bket{\sigma_|z|} = \frac{1}{2}e^{  -(t/T_2)^2  -\bar{n} (1-\cos(\chi_|qs| t) )} (\cos(\bar{n} \sin(\chi_|qs| t) + \Delta t) -1)\, ,  
\end{equation}
where $\bar{n}$ is the average photon number in the storage mode,  $\Delta$ is the pulse detuning from the qubit  frequency, and $1/T_2$ is the incoherent dephasing rate. 

\section{Simulation of qubit\textendash{cavity} coupling }
\label{sec:SimulMethodology}

We numerically simulated the qubit design shown in Fig.~2a for various qubit-position angles $\theta$ using HFSS, and for each simulation, we extracted the effective coupling rate $g$.
We treated the Josephson junction as a lumped, linear inductor in each HFSS eigenmode simulation and found the linearized mode frequencies to construct the linearized system Hamiltonian~\cite{Nigg2012}.
To treat the perturbing effect of the non-linear Josephson terms in the full Hamiltonian, we first calculated their magnitude using the energy participation-ratio method~\footnote{Z. Minev et al., in preparation}, which is based on the fields already found in the eigenmode simulation.
Second, we numerically diagonalized the full Hamiltonian to find the energy spectrum of the system.
From the spectrum, we extracted the frequencies and Kerr coefficients of the transmon and the resonator modes; from these, we calculated the coupling rate $g$.

\begin{figure}[ht!]
\includegraphics[width=3.375in]{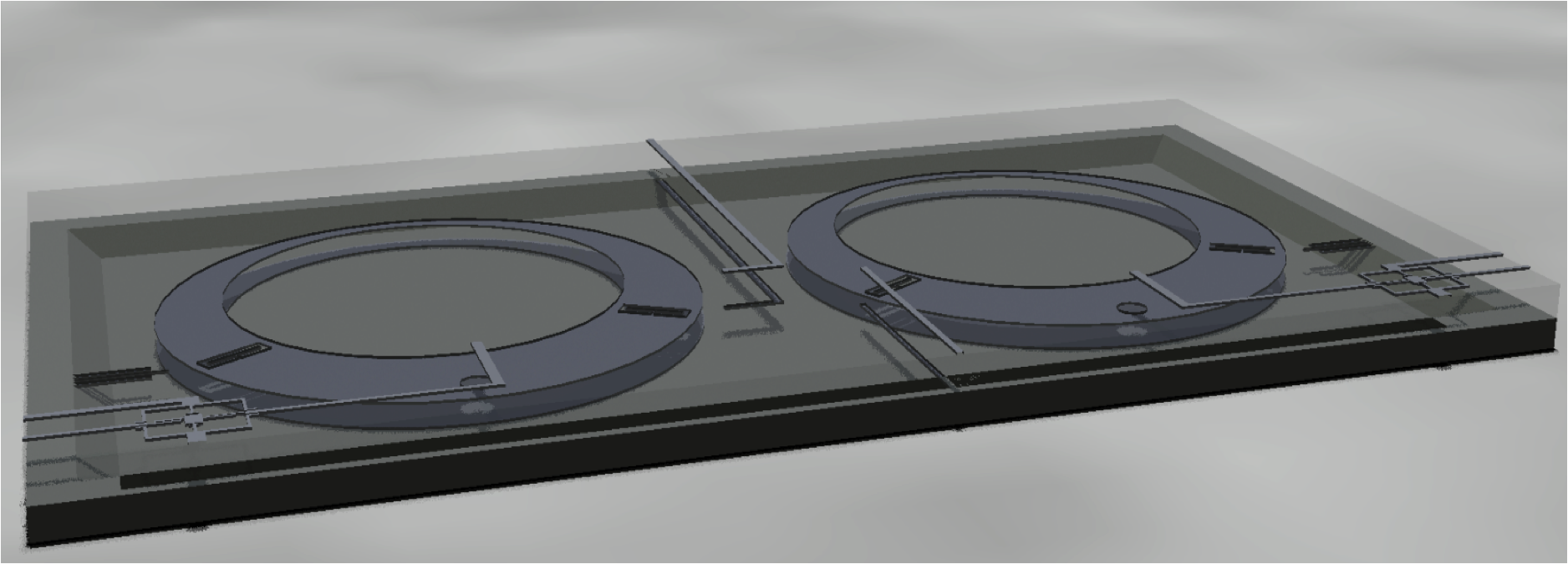}
\caption{
Sketch of a two-resonator, four-qubit device, with on-chip control lines, and readout amplifiers. The device consists of two stacked (etched) chips with 3 layers. The top layer (on the back of the top chip) houses the control lines and amplifiers (JPC type on front left and right corners). On the lower side of the top chip is a resonator-qubit layer. Apertures are used to define coupling to the control lines. The layer on the lower chip is used as the second resonator layer.}
\label{fig:vision}
\end{figure}

\bibliography{biblio}
\end{document}